\documentclass[12pt]{article}

\usepackage{color,epsfig,multicol,graphicx,axodraw,longtable}
\usepackage{amsmath,amssymb}
\newcommand{\Dlr}{\stackrel{\leftrightarrow}{D}}
\newcommand{\Dl}{\stackrel{\leftarrow}{D}}
\newcommand{\Dr}{\stackrel{\rightarrow}{D}}
\newcommand{\psl}{\not{\hspace{-0.08cm}p}}
\newcommand{\gRcf}{\frac{g_R^2 \, C_F}{16 \pi^2} }
\newcommand{\gR}{g_{\mathrm R}}
\newcommand{\HG}{H(4)}
\newcommand{\cO}{\mathcal {O}}
\newcommand{\cI}{\mathcal {I}}
\newcommand{\cF}{\mathcal {F}}
\newcommand{\cK}{\mathcal {K}}
\newcommand{\Dd}[1]{\overset{\leftrightarrow}{D}_{#1}}
\newcommand{\MS}{{\overline{\mbox{MS}}}}
\newcommand{\itep}
{~\vspace{-1.2cm}
\begin{flushright}
{\normalsize DESY 04-197}
\\
{\normalsize Edinburgh 2004/25}
\\
{\normalsize Leipzig LU-ITP 2004/023}
\\
{\normalsize Liverpool LTH 637}
\end{flushright}
\vspace{0.0cm}}
\begin{document}

\begin{center}
\itep

\begin{center}
{\Large\bf
Perturbative Renormalisation of the Second
\\ \vspace{2mm}
Moment of Generalised Parton Distributions
}
\end{center}
\vspace*{0.4cm}

{\large
M.~G\"ockeler$^{1,2}$,
R.~Horsley$^3$,
H.~Perlt$^{2,1}$,
P.~E.~L.~Rakow$^4$,
A.~Sch\"afer$^2$,
\\
\vspace{2mm}
G.~Schierholz$^{5,6}$ and
A.~Schiller$^1$
}

\vspace*{0.4cm}

{\sl
$^1$ Institut f\"ur Theoretische Physik, Universit\"at
Leipzig, \\ D-04109 Leipzig, Germany\\
$^2$ Institut f\"ur Theoretische Physik, Universit\"at
                          Regensburg,\\ D-93040 Regensburg, Germany \\
$^3$ School of Physics, University of Edinburgh, Edinburgh EH9 3JZ, UK \\
$^4$ Theoretical Physics Division, Department of Mathematical Sciences, \\
          University of Liverpool, Liverpool L69 3BX, UK \\
$^5$ John von Neumann-Institut f\"ur Computing NIC, \\
         Deutsches Elektronen-Synchrotron DESY, D-15738 Zeuthen, Germany \\
$^6$ Deutsches Elektronen-Synchrotron DESY, D-22603 Hamburg, Germany
}
\vspace*{0.4cm}

\end{center}
\begin{abstract}
  We calculate the non-forward quark matrix elements of operators with two
  covariant derivatives needed for the renormalisation of the second moment
  of generalised parton distributions in one-loop lattice perturbation theory
  using Wilson fermions. For some representations of the hypercubic group
  commonly used in simulations we determine the sets of all possible mixing
  operators. For those representations the one-loop mixing matrices of
  renormalisation factors are found. Due to non-vanishing contributions of
  operators with external ordinary derivatives the number of contributing
  operators increases compared to forward matrix elements.
\end{abstract}

\section{Introduction}

In recent years generalised parton distributions (GPDs) \cite{DM}
became a focus of both experimental and theoretical studies in hadron physics.
For an extensive up-to-date review including a comprehensive list
of references see \cite{Diehl}. GPDs provide a universal (with the
meaning used in factorisation proofs), unifying  parametrisation
for a large class of hadronic correlators, including e.g.\ form factors
and the ordinary parton distribution functions. Thus they provide a
solid formal basis
to connect information from various inclusive, semi-inclusive and
exclusive reactions in an efficient, unambiguous manner. Furthermore
they give access to physical quantities which cannot be directly
determined in experiments, like e.g.\ the orbital angular momentum
of quarks and gluons in a nucleon (for a chosen specific scheme) and
the spatial distribution of the energy or spin density of a fast
moving hadron in the transverse plane. This enormous potential motivates
the ongoing dedicated investigation of exclusive reactions at
DESY, CERN, JLab and other accelerator centers \cite{exp, DVCS}.
As GPDs are well-defined QCD objects it was possible to derive  many
fundamental theoretical results, e.g.\ the form of
their NLO-$Q^2$-evolution equations as well as the NLO coefficient
functions for Deeply Virtual Compton Scattering.

However, the direct experimental access to GPDs beyond the limiting cases
of distribution functions and simple form factors is limited.
So one can at most hope to compare experimental data with suitable
parametrisations or models. But even
under the most optimistic assumptions one would certainly need
of the order of 20 parameters per flavour for a reliable fit of all GPDs.
(Basically, because GPDs contain so much physics they cannot be expected
to be trivial functions.) Another practical problem is that exclusive
cross sections typically fall so rapidly with $Q^2$ that only moderate
$Q^2$ values can be studied. For these, however, higher-twist
contributions can be sizeable, which complicates the situation even
further.

Therefore, although there is in principle an enormous number of
reaction channels which provide information on GPDs, in practice
it is indispensable to obtain complementary information, e.g.\ from lattice
QCD calculations. On the lattice we can compute matrix elements of
local composite operators, and moments of GPDs can be related
to such matrix elements taken between states of different nucleon momenta
and spins.

More precisely we can write~\cite{ji} for example
\begin{eqnarray}
  \langle p' | \cO_{\mu_1\cdots\mu_n} |p \rangle &=& \bar u (p')
  \gamma_{(\mu_1} u(p) \sum_{i=0}^{\left[\frac{n-1}{2}\right]}
   A_{n,2i}(t)\Delta_{\mu_2}\cdots\Delta_{\mu_{2i+1}}\overline{p}_{\mu_{2i+2}}
  \cdots\overline{p}_{\mu_{n})}
  \nonumber\\
  & -&  \frac{1}{2 M}  \bar u (p') i \Delta^\alpha \sigma_{\alpha (\mu_1}
  u(p) \sum_{i=0}^{\left[\frac{n-1}{2}\right]}
  B_{n,2i}(t)\Delta_{\mu_2}\cdots\Delta_{\mu_{2i+1}}
  \overline{p}_{\mu_{2i+2}}\cdots\overline{p}_{\mu_{n})}
  \label{OpMatEl}
  \\
  & +&  C_{n}(t) {\rm Mod}(n+1,2)\frac{1}{M} \bar u (p') u(p)
  \Delta_{(\mu_1}\cdots\Delta_{\mu_n)}\, ,
  \nonumber
\end{eqnarray}
where all indices are symmetrised and trace terms are subtracted
as indicated by $(\cdots )$.
The leading twist-two operators used in (\ref{OpMatEl}) are
\begin{equation}
  \cO_{\mu_1\cdots\mu_n} = \left(\frac{i}{2}\right)^{n-1}\, \bar{\psi}
   \gamma_{( \mu_1} \Dlr_{\mu_2} \cdots  \Dlr_{\mu_n )} \psi
  \label{Opnospin}
\end{equation}
with the symmetric covariant derivative
\begin{equation}
  \Dlr \, = \Dr \, - \Dl \,.
  \label{DlDr}
\end{equation}
Analogous equations exist for
\begin{equation}
  \cO^{5}_{\mu_1\cdots\mu_n} = \left(\frac{i}{2}\right)^{n-1}\, \bar{\psi}
  \gamma_{( \mu_1} \Dlr_{\mu_2} \cdots  \Dlr_{\mu_n )} \gamma_5 \psi
  \,
  \label{Opspin}
\end{equation}
and for the tower of operators involving $\sigma_{\mu \nu}$ (related
to the generalised transversity)~\cite{hagler}.
In~(\ref{OpMatEl}) we have the nucleon 4-momentum transfer $\Delta=p'-p$
and its invariant $t=\Delta^2$, $\overline{p}=(p'+p)/2$ denotes the average
nucleon momentum and $M$ its mass.

The generalised form factors $A_{n,2i}(t)$, $B_{n,2i}(t)$ and $C_{n}(t)$
are related to the moments of the GPDs by\footnote{
As an example we give here the off-forward parton distributions
$H(x,\xi,t)$ and $E(x,\xi,t)$, one of which
reduces to the ordinary quark distribution
in the limit $\xi\to 0$ and $t\to 0$: $H(x,0,0)=q(x)$.
}
\begin{eqnarray}
  \int_{-1}^1 dx \, x^{n-1}H(x,\xi,t) &=&
  \sum_{i=0}^{\left[\frac{n-1}{2}\right]}A_{n,2i}(t)(-2\xi)^{2i}
  + {\rm   Mod}(n+1,2)C_{n}(t)(-2\xi)^{n}\, ,
  \nonumber \\
  \int_{-1}^1 dx \, x^{n-1}E(x,\xi,t) &=&
  \sum_{i=0}^{\left[\frac{n-1}{2}\right]}B_{n,2i}(t)(-2\xi)^{2i} - {\rm
  Mod}(n+1,2)C_{n}(t)(-2\xi)^{n}\, .
  \label{GPDMom}
\end{eqnarray}
The variable $\xi=-n\cdot \Delta /2$ is defined with the help of a
light-like vector $n^\mu$ which obeys $\overline{p}\cdot n = 1$.

First results for moments of GPDs obtained on the lattice were published
recently~\cite{QCDSF,Hagler:2003jd}
(see also~\cite{Gockeler:2004vx}) and soon results from
improved calculations should become available. Not surprisingly,
a number of theoretical problems still has to be settled. One
urgent task is to obtain the missing renormalisation factors and
to completely analyse the operator mixing.
The renormalisation of the operators which are related to (generalised)
parton distributions has been discussed extensively, both in the
continuum and on the lattice. Up to now, however, almost exclusively
the case of forward matrix elements has been considered. When non-forward
matrix elements are studied, new features arise, which make a
reconsideration of the renormalisation problem necessary. In particular,
the mixing with ``external ordinary derivatives'', i.e.\ with operators
of the form
$\partial_\mu \partial_\nu \cdots \left( \bar{\psi} \cdots \psi \right)$,
needs to be
investigated~\cite{shifman}. They do not contribute in forward matrix
elements, but have to be taken into account when calculating e.g.\ the
generalised form factors $A_{n,2i} (t)$ for $i > 0$.

On the lattice the mixing patterns are usually more complicated than
in the continuum, because covariance under the hypercubic group $\HG$
imposes less stringent restrictions than $O(4)$ covariance. The necessity
to consider also operators with external ordinary derivatives enlarges
the set of contributing operators even further.
These complications do not yet arise for $n=1$ and $n=2$. Hence for these
moments the renormalisation factors can be taken over from the forward case.

In this paper we investigate the renormalisation problem for $n=3$ within
the framework of one-loop lattice perturbation theory.
Some first results have been presented recently~\cite{ourlat04}.
We find that the numerical values of the
renormalisation factors are not very different from one.
We also find that the pattern of operator mixing is far more involved than
e.g.\ for moments of the ordinary parton distributions. While the
numerical results
from one-loop lattice perturbation theory have limited precision,
the results concerning the mixing of operators are valid in general.
Note also that our considerations apply equally well to moments of
distribution amplitudes.

Let us fix the notations used in our perturbative calculations.
We work in Euclidean space and use the Wilson gauge action and
Wilson fermions such that the total action is given by
$$
  S^{\rm latt}_{\rm W} = S_{\rm W,F} +S_{\rm W,G}\,.
$$
The fermionic part $S_{\rm W,F}$ for dimensionful massless
fermion fields $\psi(x)$ has the form
\begin{eqnarray}
  S_{\rm W,F}&=&
  4 a^3 r \sum_{x} \bar{\psi}(x)\psi(x)
  \nonumber
  \\
  &-& \frac{a^3}{2}
  \sum_{x,\mu}\left[\bar{\psi}(x)(r-\gamma_\mu)U_{x,\mu} \psi(x+a \hat{\mu})
  + \bar{\psi}(x+a \hat{\mu})(r+\gamma_\mu)U^\dagger_{x,\mu} \psi(x)\right]
   \,,
  \nonumber
\nonumber
\end{eqnarray}
where $a$ is the lattice spacing and the sums run over all lattice
sites $x$ and directions $\mu$
on the lattice (all other indices are
suppressed). The link matrices $U_{x,\mu}$ are related to the gauge field $A_\mu(x)$ by
$$
  U_{x,\mu} = \exp \left[i g a A_\mu(x)\right],  \quad A_\mu(x)=T^c A^c_\mu(x)\,,
$$
where $g$ is the bare gauge coupling and the $T^c$ are
the generators of the $SU(3)$ algebra.
The Wilson parameter $r$ can be chosen from
the interval $(0,1]$. The gauge action for the gluon field $A_\mu(x)$ is
\begin{equation}
  S_{\rm W,G} = \frac{6}{g^2} \sum _{x,\mu<\nu}\left[1 -
  \frac{1}{6}{\rm Tr}\left(U_{x,\mu\nu}+U^\dagger_{x,\mu\nu}\right)\right]
  \nonumber
\end{equation}
with
$$
  U_{x,\mu\nu} = U_{x,\mu}U_{x+a\hat{\mu},\nu}U^\dagger_{x+a\hat{\nu},\mu}
  U^\dagger_{x,\nu}
  \,.
$$

In the perturbative calculation the investigated operators are
sandwiched between off-shell quark states.
We shall denote the momentum of the incoming quark by $p$ and
that of the outgoing quark by $p'$.
Our calculations are performed in Feynman gauge, the final numbers
will be presented for the Wilson parameter $r=1$.

\section{Operators and mixing}
\label{OpMix}
\subsection{Renormalisation and mixing in the one-loop approximation}
\label{OpMix1}

In this section we discuss renormalisation and mixing in general
terms, on the lattice as well as in the continuum.

Let $\Gamma_j^D (p',p,\mu,\gR,\epsilon)$ ($j=1,2,\ldots,N$) be the
dimensionally regularised amputated vertex functions of $N$ mixing operators
$\cO_j$ of the same dimension
calculated in $4-2 \epsilon$ dimensions. The corresponding Born terms
are denoted by $\Gamma_j^{\mathrm {Born}}(p',p)$.
The operators potentially contributing to the mixing have to satisfy
certain symmetry requirements.
They should transform identically according to a given irreducible
representation of  $O(4)$ or $\HG$, respectively, and they should have
the same charge conjugation parity.

The (dimensionless)
renormalised coupling constant $\gR$ is related to the (dimensionful)
bare coupling constant $g$ by
\begin{equation}
  \gR^2 = \mu ^{-2\epsilon} g^2 \left( 1 + O(g^2) \right) \,,
\end{equation}
where $\mu$ is the renormalisation scale.
In one-loop perturbation theory we get results of the form
\begin{equation}
  \label{drvf}
  \begin{array}{l} \displaystyle
  \Gamma_j^D (p',p,\mu,\gR,\epsilon) = \Gamma_j^{\mathrm {Born}}(p',p)
  \\ \displaystyle \quad {}
  + \gR^2 \left[ \sum_{k=1}^N \gamma_{jk}^V \left( \frac{1}{\epsilon}
  - \gamma_E + \ln (4 \pi) - \ln \frac{(p'+p)^2}{4 \mu^2} \right)
   \Gamma_k^{\mathrm {Born}}(p',p) + f_j (p',p) \right] + O(\gR^4) \,,
  \end{array}
\end{equation}
where
$\gamma_E=0.5772\dots$ is Euler's constant.
As usual, contributions which
vanish for $\epsilon \to 0$ have been omitted.
In what follows, we systematically omit all contributions $O(\gR^4)$.
In the $\MS$
scheme the renormalised vertex functions are then given by
\begin{equation}
  \label{msvf}
  \begin{array}{l} \displaystyle
   \Gamma_j^R (p',p,\mu,\gR) = \Gamma_j^{\mathrm {Born}}(p',p)
   \\
   \displaystyle \quad {}
   + \gR^2 \left[ \sum_{k=1}^N \gamma_{jk}^V \cdot (-1) \cdot
   \ln \frac{(p'+p)^2}{4 \mu^2}
   \Gamma_k^{\mathrm {Born}}(p',p) + f_j (p',p) \right]
  \,.
  \end{array}
\end{equation}

In the absence of mixing with lower-dimensional operators the
vertex functions regularised on a lattice
can be written as
\begin{equation}
  \label{latvf}
  \begin{array}{l} \displaystyle
  \Gamma_j^L (p',p,a,\gR) = \Gamma_j^{\mathrm {Born}}(p',p)
  \\ \displaystyle \quad {}
  + \gR^2 \left[ \sum_{k=1}^N \gamma_{jk}^V \cdot (-1) \cdot
  \ln \frac{ a^2 (p'+p)^2}{4} \, \Gamma_k^{\mathrm {Born}}(p',p)
  + f_j^L (p',p) \right]
  \end{array}
\end{equation}
up to terms vanishing as $a \to 0$.
There should be an $N \times N$ matrix $\zeta$ such that the relation
between the bare lattice vertex functions and the $\MS$
renormalised vertex functions can be written as
\begin{equation}
  \label{zeta}
  \Gamma_j^R (p',p,\mu,\gR) =
  \sum_{k=1}^N \left( \delta_{jk} + \gR^2 \zeta_{jk}
  \right)
  \Gamma_k^L (p',p,a,\gR) \,.
\end{equation}
So we should have
\begin{equation}
  \begin{array}{l} \displaystyle
  \Gamma_j^R (p',p,\mu,\gR) = \Gamma_j^{\mathrm {Born}}(p',p)
  \\ \displaystyle \quad {}
  + \gR^2 \left[ \sum_{k=1}^N \left(\zeta_{jk} - \gamma_{jk}^V
  \ln \frac{a^2 (p'+p)^2}{4} \right) \,
    \Gamma_k^{\mathrm {Born}}(p',p) + f_j^L (p',p) \right]
  \,.
\end{array}
\end{equation}
Comparing with (\ref{msvf}) we arrive at
\begin{equation}
  \sum_{k=1}^N \left(\zeta_{jk} - \gamma_{jk}^V
  \ln \left(  a^2 \mu^2 \right) \right) \, \Gamma_k^{\mathrm {Born}}(p',p)
  + f_j^L (p',p) - f_j (p',p) = 0 \,.
\end{equation}
As this equation must hold for arbitrary momenta $p'$, $p$, there should
be constants $c_{jk}^V$ such that
\begin{equation}
  \label{const}
  f_j^L (p',p) - f_j (p',p) =
  \sum_{k=1}^N c_{jk}^V \, \Gamma_k^{\mathrm {Born}}(p',p) \,.
\end{equation}
This fixes the matrix $\zeta_{jk}$:
\begin{equation}
  \label{zeta2}
  \zeta_{jk} = \gamma_{jk}^V \ln \left( a^2 \mu^2 \right) - c_{jk}^V \,.
\end{equation}
If the coefficients $c_{jk}^V$ are uniquely determined, exactly all $N$
operators mix to one-loop accuracy.
It might happen that certain operators contribute to mixing only in higher orders.
In that case $N$ is decreased.
If Eq.~(\ref{const}) cannot be satisfied, this shows that at least one
mixing operator has been overlooked.

Mixing with lower-dimensional operators leads to the appearance of
additional terms on the r.h.s.\ of Eq.~(\ref{latvf}). For example, consider
the case of a single operator which mixes with $\cO_j$ and
whose dimension is one unit smaller. (Such a case
will appear in our applications.) Then we get instead of (\ref{latvf})
\begin{equation}
  \begin{array}{l} \displaystyle
  \Gamma_j^L (p',p,a,\gR) = \Gamma_j^{\mathrm {Born}}(p',p)
  \\ \displaystyle
  \ \
  + \gR^2 \left[ \sum_{k=1}^N \gamma_{jk}^V \cdot (-1) \cdot
  \ln \frac {a^2 (p'+p)^2}{4} \, \Gamma_k^{\mathrm {Born}}(p',p)
  + \frac{1}{a} c_j \, \Gamma^{\mathrm {Born}}(p',p)
  + f_j^L (p',p) \right]
  \,,
  \end{array}
\end{equation}
where $\Gamma^{\mathrm {Born}}(p',p)$ is the Born term of the additional
lower-dimensional
operator. The $1/a$ contribution has to be subtracted from
$\Gamma_j^L (p',p,a,\gR)$ before the connection with
$\Gamma_j^R (p',p,\mu,\gR)$ can be established, i.e.\ in (\ref{zeta})
$\Gamma_k^L (p',p,a,\gR)$ has to be replaced by
\begin{equation}
  \label{subtr}
  \Gamma_k^L (p',p,a,\gR) - \frac{\gR^2}{a} c_k \, \Gamma^{\mathrm {Born}}(p',p) \,.
\end{equation}
Then Eq.~(\ref{const}) is obtained as before.

In order to compute the matrix $Z_{jk}$ of renormalisation and
mixing coefficients we note that
the connection between the bare lattice vertex
functions and the $\MS$ renormalised vertex functions
can be written as
\begin{equation}
  \Gamma_j^R (p',p,\mu,\gR) = Z_\psi^{-1}
  \sum_{k=1}^N Z_{jk} \, \Gamma_k^L (p',p,a,\gR)
\end{equation}
with the quark wave function renormalisation constant%
\footnote{Often the quark wave function renormalisation constant
is defined as $1/Z_\psi$.}
$Z_\psi$.
(In the presence of mixing with a lower-dimensional operator,
$\Gamma_k^L (p',p,a,\gR)$ has again to be replaced by the subtracted
expression (\ref{subtr})).
Comparison with (\ref{zeta}) yields
\begin{equation}
  \label{zz}
  Z_\psi^{-1} Z_{jk} = \delta_{jk} + \gR^2 \zeta_{jk}
  \,.
\end{equation}

The quark wave function renormalisation constant is calculated from
the quark propagator. Write the lattice regularised inverse quark
propagator in Feynman gauge
(after subtracting a linearly diverging contribution $\sim 1/a$) as
\begin{equation}
  S_L^{-1} (p) = i  \psl  \left[ 1 +
 \gRcf
 \left( -\ln \left(a^2 p^2 \right) - \sigma_L \right)
  \right]
\end{equation}
where $\sigma_L=11.8524$ (see e.g.~\cite{Capitani:2001xi}) for Wilson fermions.
In dimensional regularisation the inverse quark propagator is given by
\begin{equation}
S_D^{-1} (p) =i \psl \left[ 1 +
  \gRcf
  \left( \frac{1}{\epsilon} - \gamma_E + \ln (4 \pi)
  - \ln \frac{ p^2}{\mu^2} + 1 \right)
  \right]
\end{equation}
and we get for the $\MS$ renormalised propagator
\begin{equation}
  S_R^{-1} (p) = i \, p\hspace{-1.1ex}/ \left[ 1 + \gRcf
   \left( - \ln  \frac{p^2}{\mu^2} +1 \right)
  \right] \,.
\end{equation}
Defining $Z_\psi$ such that
\begin{equation}
  S_R^{-1} (p) = Z_\psi^{-1} S_L^{-1} (p)
\end{equation}
we have
\begin{equation}
  Z_\psi = 1 - \gRcf
  \left( \ln \left( a^2 \mu^2 \right) + 1 + \sigma_L \right)
   \,.
\end{equation}
With the help of this result we get from (\ref{zeta2}) and (\ref{zz})
for the matrix of the renormalisation and mixing coefficients
\begin{equation}
  \label{zjk}
  Z_{jk} = \delta_{jk} + \gR^2 \left[
  \left( \gamma_{jk}^V - \delta_{jk} \frac{C_F}{16 \pi^2} \right)
  \ln \left( a^2 \mu^2 \right) - c_{jk}^V -
  \delta_{jk} \frac{C_F}{16 \pi^2} (1 + \sigma_L) \right]
\,.
\end{equation}

The basic computational task is thus to find the functions $f_j (p',p)$ and
$f_j^L (p',p)$ in Eqs.~(\ref{drvf}) and (\ref{latvf}) (or rather their
difference). Then we can compute the coefficients $c_{jk}^V$ from
Eq.~(\ref{const}) and use Eq.~(\ref{zjk}) to get the desired
renormalisation and mixing coefficients.
If we are only interested in the renormalisation of one particular
operator, corresponding to $j=1$ say, it is sufficient to restrict the
calculations to the case $j=1$.

\subsection{Contributing operators for special representations}
\label{OpMix2}

Let us introduce the self-explaining notations for operators with
covariant and external ordinary derivatives
\begin{eqnarray}
  \cO_{\mu\nu\omega}^{DD}&=& -\frac{1}{4}
  \bar\psi \gamma_\mu \Dlr_\nu \Dlr_\omega\psi\,,
  \nonumber
  \\
  \cO_{\mu\nu\omega}^{\partial D}&=& -\frac{1}{4}
  \partial_\nu \left( \bar\psi \gamma_\mu \Dlr_\omega\psi \right) \,,
  \label{OpDD}
  \\
  \cO_{\mu\nu\omega}^{\partial \partial}&=& -\frac{1}{4}
  \partial_\nu \partial_\omega
  \left(\bar\psi \gamma_\mu \psi \right)
  \nonumber
\end{eqnarray}
and
\begin{eqnarray}
  \cO_{\mu\nu\omega}^{5,DD}&=&-\frac{1}{4}
  \bar\psi \gamma_\mu \gamma_5 \Dlr_\nu \Dlr_\omega\psi\,,
  \nonumber
  \\
  \cO_{\mu\nu\omega}^{5,\partial D}&=& -\frac{1}{4}
  \partial_\nu \left( \bar\psi \gamma_\mu \gamma_5 \Dlr_\omega \psi\right) \,,
  \label{Op5DD}
  \\
  \cO_{\mu\nu\omega}^{5,\partial \partial}&=&-\frac{1}{4}
  \partial_\nu \partial_\omega
  \left(\bar\psi \gamma_\mu \gamma_5\psi \right) \,,
  \nonumber
\end{eqnarray}
as well as the lower-dimensional operators
\begin{equation}
  \cO_{\mu\nu\omega}^D=-\frac{i}{2}
  \bar\psi [ \gamma_\mu , \gamma_\nu ] \Dlr_\omega \psi \,,
  \quad
  \cO_{\mu\nu\omega}^\partial=-\frac{i}{2}
  \partial_\omega
  \left( \bar\psi  [\gamma_\mu,\gamma_\nu ]  \psi \right)\,.
  \label{Oplower}
\end{equation}
For completeness we include here also operators contributing to
non-forward {\it transversity} matrix elements with two
derivatives which were not considered in Eqs.~(\ref{Opnospin})
and (\ref{Opspin}):
\begin{equation}
  \cO_{\mu\nu\omega\sigma}^{T,DD}=-\frac{1}{4}\bar\psi [\gamma_\mu,\gamma_\nu]
  \Dlr_\omega   \Dlr_\sigma\psi\,,
  \quad
  \cO_{\mu\nu\omega\sigma}^{T,\partial\partial }=
  -\frac{1}{4} \partial_\omega\partial_\sigma
  \left( \bar\psi [\gamma_\mu,\gamma_\nu] \psi  \right) \,.
  \label{Optrans}
\end{equation}
As short-hand notations we use in the following (cf.~\cite{group})
\begin{eqnarray}
  \cO_{ \{ \nu_1\nu_2\nu_3 \} }&=& \frac{1}{6} \left(
  \cO_{\nu_1\nu_2\nu_3}+\cO_{\nu_1\nu_3\nu_2}+\cO_{\nu_2\nu_1\nu_3} +
  \cO_{\nu_2\nu_3\nu_1}+\cO_{\nu_3\nu_1\nu_2}+\cO_{\nu_3\nu_2\nu_1}  \right)
  \,,
  \\
  \cO_{\|\nu_1\nu_2\nu_3\| } &=& \cO_{\nu_1\nu_2\nu_3}-\cO_{\nu_1\nu_3\nu_2}+
  \cO_{\nu_3\nu_1\nu_2}-\cO_{\nu_3\nu_2\nu_1}-2\,\cO_{\nu_2\nu_3\nu_1}
  +2\,\cO_{\nu_2\nu_1\nu_3}
  \,,
  \\
  \cO_{\langle\langle\nu_1\nu_2\nu_3\rangle\rangle } &=&
  \cO_{\nu_1\nu_2\nu_3}+\cO_{\nu_1\nu_3\nu_2}
  -\cO_{\nu_3\nu_1\nu_2}-\cO_{\nu_3\nu_2\nu_1}
  \,.
\end{eqnarray}

First we consider the operator
\begin{equation}
  \label{O11}
  \cO_{\{124\}}^{DD}
   = - \frac{1}{4} \bar{\psi} \gamma_{\{1} \Dd{2}\Dd{4\}} \psi \,.
\end{equation}
Its charge conjugation parity is $C=-1$ and it is
a member of an irreducible multiplet of operators transforming
according to the representation $\tau^{(4)}_2$ of $\HG$~\cite{group}.
Here $\tau_k^{(l)}$ denotes an irreducible representation of $\HG$
with dimension $l$,
and $k=1,2,\dots$ labels inequivalent representations of the same dimension.
This operator can only mix with
\begin{equation}
  \label{O11dd}
   \cO_{\{124\}}^{\partial\partial}= - \frac{1}{4} \partial_{\{2} \partial_{4}
   \left(\bar{\psi} \gamma_{1\}}\psi \right) \,.
\end{equation}

Next we examine the operator
\begin{equation}
  \cO_1=\cO^{DD}_{\{114\}}-\frac{1}{2}
  \left(\cO^{DD}_{\{224\}}+\cO^{DD}_{\{334\}}\right)
  \,.
  \label{O1}
\end{equation}
It has already been used in lattice computations of forward hadronic matrix
elements, because in this case it suffers
only from rather mild mixing problems.
It belongs to the representation $\tau^{(8)}_1$ with $C=-1$.

Taking into account also external ordinary derivatives, one finds the following
operators which transform identically and could
therefore mix with (\ref{O1})\footnote
{The charge conjugation parities $C$ of $\cO_7$ and $\cO_5$ coincide since 
$\cO_7$ is antisymmetric in the indices of the covariant derivatives.}:
\begin{eqnarray}
  &&\cO_2=\cO^{\partial\partial}_{\{114\}}-\frac{1}{2} \left(
   \cO^{\partial\partial}_{\{224\}}
  +\cO^{\partial\partial}_{\{334\}}\right)
  \,,
  \nonumber
  \\
  &&\cO_3=\cO^{DD}_{\langle\langle 114\rangle\rangle}-\frac{1}{2}
  \left( \cO^{DD}_{\langle\langle224\rangle\rangle}+
         \cO^{DD}_{\langle\langle334\rangle\rangle}\right)
  \,,
  \nonumber
  \\
  &&\cO_4=\cO^{\partial\partial}_{\langle\langle 114 \rangle\rangle}-\frac{1}{2}
  \left(  \cO^{\partial\partial}_{\langle\langle 224 \rangle\rangle}+
          \cO^{\partial\partial}_{\langle\langle 334 \rangle\rangle}\right)
  \,,
  \label{O4}
  \\
  &&\cO_5=\cO^{5,\partial D}_{||213||}
  \,,
  \nonumber
  \\
  &&\cO_6=\cO^{5,\partial D}_{\langle\langle213\rangle\rangle}
  \,,
  \nonumber
  \\
  &&\cO_7=\cO^{5,DD}_{||213||}
  \nonumber
\end{eqnarray}
and the lower-dimensional operator
\begin{equation}
  \label{O8}
  \cO_8=   \cO^\partial_{411}-
  \frac{1}{2}\left(\cO^\partial_{422}+\cO^\partial_{433} \right)
  \,.
\end{equation}

As an ``axial'' analogue of (\ref{O11}) we consider the operator
$\cO^{5,DD}_{\{124\}}$
\begin{equation}
  \label{O511}
  \cO_{\{124\}}^{5,DD}= -\frac{1}{4}\bar{\psi} \gamma_{\{1} \Dd{2}\Dd{4\}}
  \gamma_5\psi\,,
\end{equation}
which can mix with
\begin{equation}
  \label{O511dd}
   \cO_{\{124\}}^{5,\partial\partial}=-\frac{1}{4} \partial_{\{2} \partial_{4}
  \left(\bar{\psi}    \gamma_{1\}} \gamma_5\psi \right)\, .
\end{equation}
(\ref{O511}) and (\ref{O511dd}) belong to $\tau_3^{(4)}$ with $C=+1$.

Similarly, we have as a counterpart of (\ref{O1}) the operator
\begin{equation}
  \cO^5_1=\cO^{5,DD}_{\{114\}}-\frac{1}{2}
  \left(\cO^{5,DD}_{\{224\}}+\cO^{5,DD}_{\{334\}}\right)
  \,.
\label{O51}
\end{equation}
Its charge conjugation parity is $C=+1$ and it is
a member of an irreducible multiplet of operators transforming
according the representation $\tau^{(8)}_2$ of $\HG$~\cite{group}.

The following operators with identical transformation behaviour could
potentially mix with (\ref{O51}):
\begin{eqnarray}
  &&\cO^5_2=\cO^{5,\partial\partial}_{\{114\}}-\frac{1}{2} \left(
  \cO^{5,\partial\partial}_{\{224\}}
  +\cO^{5,\partial\partial}_{\{334\}}\right)
  \,,
  \nonumber
  \\
  &&\cO^5_3=\cO^{5,DD}_{\langle\langle 114\rangle\rangle}-\frac{1}{2}
  \left( \cO^{5,DD}_{\langle\langle224\rangle\rangle}+
         \cO^{5,DD}_{\langle\langle334\rangle\rangle}\right)
  \,,
  \nonumber
  \\
  &&\cO^5_4=\cO^{5,\partial\partial}_{\langle\langle 114
  \rangle\rangle}-\frac{1}{2}
  \left(  \cO^{5,\partial\partial}_{\langle\langle 224 \rangle\rangle}+
       \cO^{5,\partial\partial}_{\langle\langle 334 \rangle\rangle}\right)
  \,,
  \label{O54}
  \\
  &&\cO^5_5=\cO^{\partial D}_{||213||}
  \,,
  \nonumber
  \\
  &&\cO^5_6=\cO^{\partial D}_{\langle\langle213\rangle\rangle}
  \,,
  \nonumber
  \\
  &&\cO^5_7=\cO^{DD}_{||213||}
  \nonumber
\end{eqnarray}
and the lower-dimensional operator
\begin{equation}
  \label{O58}
  \cO^5_8 =
  \cO^D_{123} -  2\cO^D_{231} - \cO^D_{132}
  \,.
\end{equation}

For the operators (\ref{Optrans}) we consider the representations
$\tau_2^{(3)}$, $\tau_3^{(3)}$ and $\tau_2^{(6)}$ with $C=-1$~\cite{Meinulf}.
In the case of $\tau_2^{(3)}$ we choose the
representative operator
\begin{equation}
  \cO^T_1=\cO^{T,DD}_{4\{123\}}\,,
  \label{OT1}
\end{equation}
which may mix with
\begin{equation}
  \cO^T_2=\cO^{T,\partial\partial}_{4\{123\}} \,.
  \label{OT2}
\end{equation}
For $\tau_3^{(3)}$ we take
\begin{equation}
  \cO^T_3=-\cO^{T,DD}_{1\{133\}}+\cO^{T,DD}_{1\{144\}}
          -\cO^{T,DD}_{2\{233\}}+\cO^{T,DD}_{2\{244\}}
         -2\cO^{T,DD}_{3\{344\}}
  \label{OT3}
\end{equation}
mixing with
\begin{equation}
  \cO^T_4=-\cO^{T,\partial\partial}_{1\{133\}}
          +\cO^{T,\partial\partial}_{1\{144\}}
          -\cO^{T,\partial\partial}_{2\{233\}}
          +\cO^{T,\partial\partial}_{2\{244\}}
         -2\cO^{T,\partial\partial}_{3\{344\}} \,.
  \label{OT4}
\end{equation}
The operator
\begin{equation}
  \cO^T_5 = \cO^{T,DD}_{13\{32\}} + \cO^{T,DD}_{23\{31\}}
          - \cO^{T,DD}_{14\{42\}} - \cO^{T,DD}_{24\{41\}}
  \label{OT5}
\end{equation}
belonging to the representation $\tau_2^{(6)}$ also mixes
with only one additional operator:
\begin{equation}
  \cO^T_6 =  \cO^{T,\partial\partial}_{13\{32\}}
           + \cO^{T,\partial\partial}_{23\{31\}}
           - \cO^{T,\partial\partial}_{14\{42\}}
           - \cO^{T,\partial\partial}_{24\{41\}} \,.
  \label{OT6}
\end{equation}
Other representations show a more complicated
mixing behaviour and will be not considered here.

\section{One-loop calculation of non-forward matrix elements}

\subsection{Calculational technique}

We calculate the matrix elements of the operators
in one-loop lattice perturbation theory
in the infinite volume limit following Kawai et al.~\cite{Kawai:1980ja}.
The computation is performed symbolically
adopting and significantly extending
a {\it Mathematica} program package developed originally
for the case of moments of structure functions using
Wilson~\cite{Gockeler:1996hg},
clover~\cite{Capitani:2001xi} and overlap fermions~\cite{Horsley:2004mx}.

Using that approach we have full analytic control over
pole cancellation.
The Lorentz index structure of the matrix elements
is left completely free, so that we are able to construct all
representations of the hypercubic group for the second moments
in the non-forward case.
The analytic derivation is disconnected from the numerical calculation
of {\it finite} lattice integrals.
These integrals are calculated to a high accuracy and are given as
look-up tables.

To be more precise, we recapitulate in short the
strategy~\cite{Kawai:1980ja,Gockeler:1996hg} for calculating the diagrams
adapted to our case of two distinct external momenta $p \not= p'$.
Basically we have to evaluate a typical lattice integral of the form
\begin{equation}
  \cI_{\mu_1\cdots\mu_n}(a,p',p) = \int_{-\pi/a}^{\pi/a}
  \frac{d^4 k}{(2\pi)^4}\, \cK_{\mu_1\cdots\mu_n}(a,p',p,k)
  \label{LattInt}
\end{equation}
where the integration is performed  over the first Brillouin zone.
The numerator of the kernel $\cK_{\mu_1\cdots\mu_n}(a,p',p,k)$
is a polynomial
in sines and cosines of the lattice momenta $k$, the denominator contains
the denominators of
lattice quark and gluon propagators.
Such an  integral $\cI$ is calculated by splitting it into two parts
\begin{equation}
  \cI = \tilde{\cI} + ( \cI - \tilde{\cI}) \,.
  \label{LattIntDec}
\end{equation}
Here $\tilde{\cI}$ denotes the Taylor expansion of the original
integral $\cI$ in the external momenta
\begin{eqnarray}
  \tilde{\cI}(a,p',p) &=& \cI(a,0,0)
  \nonumber
  \\
  &+& \sum_\alpha \left\{ p'_\alpha
  \frac{\partial \cI (a,p' ,p)}{\partial p^\prime_\alpha}\Big|_{p'=p=0}
  + p_\alpha
  \frac{\partial \cI (a,p' ,p)}{\partial p_\alpha}\Big|_{p'=p=0}
  \right\}
  + \dots
\end{eqnarray}
where the order of the expansion
is given by the degree of ultraviolet (UV) divergence of $\cI$.
As a consequence, the difference $ \cI - \tilde{\cI}$ is UV
finite and can be calculated in
the (Euclidean) continuum, taking the limit $a\to 0$.
The original UV poles appear now as infrared (IR) poles in the Taylor
expansion and are regularised
using dimensional regularisation with $d>4$.
Since in that case
$\tilde {\cI}|_{a\to 0}$ vanishes, we have
$ (\cI - \tilde{\cI})|_{a\to 0}=
\cI|_{a\to 0}\equiv \cI^{\rm cont}(p',p)$
and the second contribution to the final lattice result is just a
one-loop continuum calculation in dimensional regularisation.

The first (Taylor expanded) part is calculated in $d=4- 2 \epsilon$
dimensions at finite $a$, the poles in $\epsilon$ analytically cancel those
of the second part $\cI^{\rm cont}$.
Note that the first part depends on the external momenta only via the Taylor
expansion, therefore the lattice integrals (the expansion coefficients) are
just numbers (independent of $p$ and $p'$).

In the Euclidean continuum part $\cI^{\rm cont} (p',p)$ the finite
contribution depends on the (different) external
momenta in a complicated manner.
As in Minkowski space, the finite part of three-point functions with three
different propagators is difficult  to represent in a simple compact analytic
form (see, e.g., \cite{DavydychevTarasovCampbell}).
In parametrising that contribution we have chosen the following form
\begin{eqnarray}
  \cI^{\rm cont}_{\mu_1\cdots\mu_n}(p',p) = \frac{1}{\epsilon}
  A_{\mu_1\cdots\mu_n}(p',p)
  + \sum_{i,j,k,m} B_{\mu_1\cdots\mu_n} (p',p,i,j,k,m)
  \cF (i,j,k,m,p',p)\,.
  \label{ICont}
\end{eqnarray}
In (\ref{ICont}) the functions $A_{\mu_1\cdots\mu_n}(p',p)$ (known analytically)
and $B_{\mu_1\cdots\mu_n}(p',p,i,j,k,m)$
contain the general index structure assigned to the external momenta $p$
and $p'$. The symbols $\cF$ stand for the remaining integrals (finite for
$p\ne p'$) over the Feynman parameters\footnote{One Feynman parameter integration
can be performed analytically, leading, however, to lengthy
expressions for the individual integrals.
We have checked that for our continuum results we recover the known results
of the forward case.}:
\begin{eqnarray}
  \cF (i,j,k,m,p',p) = \int_0^1 dx \int_0^{1-x} dy\, x^i \,y^j
  \left(Q^2(x,y,p',p)\right)^k \ln^m \frac{Q^2(x,y,p',p)}{\mu^2}
\end{eqnarray}
with
$$
Q^2(x,y,p',p) = p^2\, x(1-x) + p^{\prime 2} \, y(1-y)  -2 \, p \cdot p' \, x\,y \,.
$$
Only few combinations of the integers  $i,j,k,m$
are actually needed in the ``continuum part'' of the one-loop calculation.
But compact expressions for the $\cF (i,j,k,m,p',p)$ do not seem to exist.

As we have shown in Section~\ref{OpMix1}, in order to calculate the
finite renormalisation matrix converting the lattice result to the $\MS$ scheme
we have to find the difference between the lattice
and the continuum one-loop contribution
$f_j^L(p',p)-f_j(p',p)$, and similarly for the wave function renormalisation.
Therefore, we do not need the explicit form of the finite continuum
one-loop contribution (which is essentially given by the $\cF$'s),
since it cancels exactly the
part coming from the second term of~(\ref{LattIntDec}).
As a consequence, we do not present $\cF$-integrals here.
However, choosing other schemes, the finite contributions to
$\cI_{\mu_1\cdots\mu_n}(p',p)^{\rm cont}$ might be needed.

\subsection{One-loop renormalisation matrices for the chosen representations}

\begin{figure}[!htb]
 \begin{center}
  \begin{picture}(400,100)(0,0)
    \SetOffset(30,-10)
    \GlueArc(100,20)(30,0,180){5}{10}
    \ArrowLine(10,20)(70,20)
    \ArrowLine(70,20)(130,20)
    \ArrowLine(130,20)(190,20)
    \Text(30,5)[bl]{$p$}
    \Text(90,5)[bl]{$p+k$}
    \Text(160,5)[bl]{$p$}
    \Text(100,75)[tl]{$k$}
    \ArrowLine(210,20)(260,20)
    \ArrowLine(260,20)(320,20)
    \GlueArc(260,45)(20,-90,270){5}{13}
    \Text(235,5)[bl]{$p$}
    \Text(285,5)[bl]{$p$}
    \Text(260,80)[bl]{$k$}
  \end{picture}
 \end{center}
\caption{One-loop diagrams contributing to the quark self-energy.}
 \label{fig:self}
\end{figure}
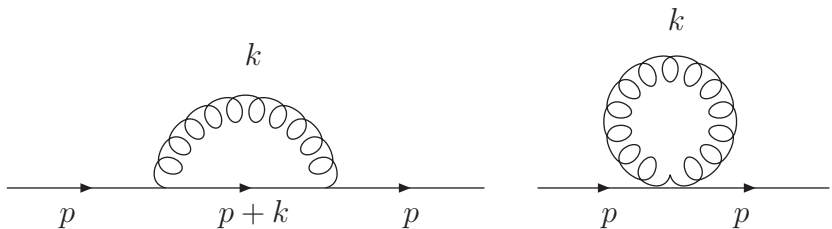

To calculate the
one-loop contributions for the above operators
between quark states we have to take into account
self energy and amputated Green function diagrams.
The quark self energy diagrams (contributing to the quark
wave function renormalisation) are shown in Fig.~\ref{fig:self}.
Denoting the operator insertions
by a black dot, the vertex and tadpole diagrams (Fig.~\ref{fig:vertex})
as well as the ``cockscomb'' diagrams (Fig.~\ref{fig:cockscomb})
contribute in one-loop order to the amputated Green functions.
\begin{figure}[!htb]
 \begin{center}
  \begin{picture}(400,130)(0,0)
   \SetOffset(30,-10)
   \ArrowLine(10,10)(36,50)
   \ArrowLine(36,50)(70,100)
   \ArrowLine(70,100)(104,50)
   \ArrowLine(104,50)(130,10)
   \GCirc(70,100){7}{0}
   \Gluon(36,50)(104,50){5}{8}
   \Text(10,30)[bl]{$p$}
   \Text(100,70)[bl]{}
   \Text(125,30)[bl]{$p'$}
   \SetOffset(230,-10)
   \ArrowLine(36,10)(70,60)
   \ArrowLine(70,60)(104,10)
   \GCirc(70,60){7}{0}
   \GlueArc(70,90)(25,-90,270){5}{13}
   \Text(30,30)[bl]{$p$}
   \Text(105,30)[bl]{$p'$}
  \end{picture}
 \end{center}
\caption{One-loop vertex and tadpole diagrams.}
 \label{fig:vertex}
\end{figure}
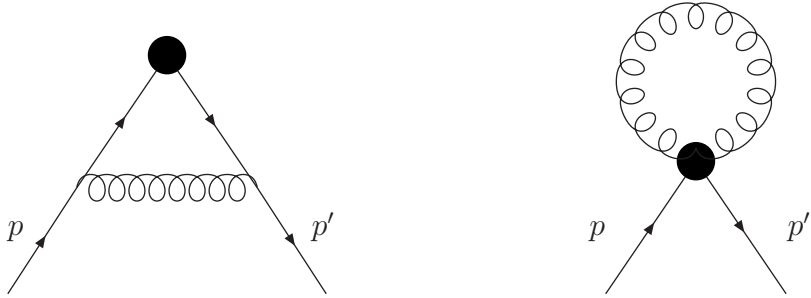
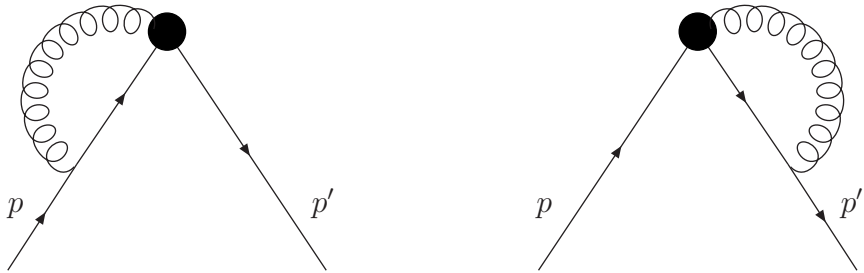
\begin{figure}[!htb]
 \begin{center}
  \begin{picture}(400,130)(0,0)
   \SetOffset(30,-10)
   \ArrowLine(10,10)(36,50)
   \ArrowLine(36,50)(70,100)
   \ArrowLine(70,100)(130,10)
   \GCirc(70,100){7}{0}
   \GlueArc(50,75)(30,60,240){5}{10}
   \Text(10,30)[bl]{$p$}
   \Text(100,70)[bl]{}
   \Text(125,30)[bl]{$p'$}
   \ArrowLine(210,10)(270,100)
   \ArrowLine(270,100)(304,50)
   \ArrowLine(304,50)(330,10)
   \GCirc(270,100){7}{0}
   \GlueArc(290,75)(30,-60,120){5}{10}
   \Text(210,30)[bl]{$p$}
   \Text(300,70)[bl]{}
   \Text(325,30)[bl]{$p'$}
  \end{picture}
 \end{center}
\caption{One-loop cockscomb diagrams.}
 \label{fig:cockscomb}
\end{figure}

We present the results for the Z factors of the
operators discussed in Section~\ref{OpMix2}
in the generic form
\begin{equation}
  Z_{jk}^{(m)} = \delta_{jk} -
  \gRcf \left(\gamma_{jk}\,\ln(a^2\mu^2)
  + c_{jk}^{(m)} \right)
\end{equation}
where (cf.\ (\ref{zjk}))
\begin{equation}
   \gamma_{jk}=\delta_{jk}- \frac{16 \pi^2}{C_F} \gamma_{jk}^V\,, \quad
  c_{jk}^{(m)}=\delta_{jk} (1+\sigma_L )+\frac{16 \pi^2}{C_F} c_{jk}^V
  \,.
\end{equation}
The superscript $(m)$ with $m=I,II$ distinguishes the
realisations I and II of the covariant derivatives, which are explained
in the Appendix.

\subsubsection{$\cO_{\{124\}}^{DD} \, (\tau_2^{(4)},C=-1)$}

For this representation we find mixing between
$\cO_{\{124\}}^{DD}$ (\ref{O11}) and
$\cO^{\partial\partial}_{\{124\}}$ (\ref{O11dd}). The
corresponding $2 \times 2$-mixing matrices are
\begin{equation}
  \gamma_{jk}=\left(
  \begin{array}{rr}
     \frac{25}{6} & -\frac{5}{6}\\
       0          & 0
  \end{array}
  \right) \,,
  \label{andim1}
\end{equation}
\begin{equation}
  c_{jk}^{(I,II)} =\left(
  \begin{array}{rr}
       -11.5632     &  0.0241\\
         0          & 20.6178
  \end{array}
  \right) \,.
  \label{cjk}
\end{equation}
The anomalous dimension matrix was calculated earlier~\cite{shifman}.
The element $c_{11}=-11.5632$ in the matrix (\ref{cjk}) is known from
our previous calculation in the forward case~\cite{Gockeler:1996hg}
and $c_{22}$ comes from the renormalisation of the local vector
current~\cite{Capitani:2001xi}.
The matrix $c_{jk}^{(I,II)}$ shows a very small mixing between the operators
$\cO_{\{124\}}^{DD}$ and $\cO^{\partial\partial}_{\{124\}}$.
Thus it may be justified to neglect the mixing in practical applications.


\subsubsection{$\cO_1 \, (\tau_1^{(8)},C=-1)$}

We consider the mixing of operators having the same dimension first.
The relevant operators
have been defined in Section~\ref{OpMix2}. To one-loop accuracy the operator
$\cO_7$ in Eq.~(\ref{O4}) does not contribute  and we have to consider the
following mixing set:
$$
  \{\cO_1,\cO_2, \cO_3,\cO_4,\cO_5,\cO_6\} \,.
$$
The
anomalous dimension matrix is
\begin{equation}
  \gamma_{jk} =\left( \begin{array}{rrrrrr}
 \frac{25}{6}&-\frac{5}{6}& 0          & 0          & 0          & 0\\
    0        & 0          & 0          & 0          & 0          & 0\\
    0        & 0          & \frac{7}{6}&-\frac{5}{6}&1           &-\frac{3}{2}\\
    0        & 0          & 0          & 0          & 0          & 0 \\
    0        & 0          & 0          & 0          & 2          & -2 \\
    0        & 0          & 0          & 0          & -\frac{2}{3}& \frac{2}{3}
  \end{array}
  \right)
  \label{andim2}
\end{equation}
and the finite  part of the mixing matrix is given by
\begin{equation}
  c_{jk}^{(I)} =\left( \begin{array}{rrrrrr}
 -12.1274      & -2.7367      & 0.3685      & 0.9934      & 0.0156   & 0.1498\\
    0          &  20.6178     & 0           & 0           & 0        & 0\\
  3.3060       &  18.1841     & -14.8516    & -4.3023     & -0.9285  & 0.7380 \\
    0          & 0            & 0           &  20.6178    & 0        & 0 \\
 0             & 3.2644       & 0           & 0           & 0.3501   & 0.0149 \\
 0             & 3.2644       & 0           & 0           & 0.0050   & 0.3600
  \end{array}
  \right)
\end{equation}
or
\begin{equation}
  c_{jk}^{(II)} =\left( \begin{array}{rrrrrr}
 -12.1274      & 1.4913       & 0.3685      & -0.4160     & 0.0156   & 0.1498\\
    0          &  20.6178     & 0           & 0           & 0        & 0\\
  3.3060       & -8.01456     & -14.8516    &  4.3023     & -0.9285  & 0.7380 \\
    0          & 0            & 0           &  20.6178    & 0        & 0 \\
 0             & 3.2644       & 0           & 0           & 0.3501   & 0.0149 \\
 0             & 3.2644       & 0           & 0           & 0.0050   & 0.3600
  \end{array}
  \right) \,.
\end{equation}
The matrices $c_{jk}^{(I,II)}$ show a sizeable mixing
of the operator $\cO_1$ with other operators, especially $\cO_2$
containing two external ordinary derivatives.
This mixing becomes irrelevant in the forward case, where the matrix element
of $\cO_2$ vanishes.
On the other hand, the
mixing between the operators $\cO_1$ and $\cO_3$ is already known from
our previous calculations in the forward case~\cite{Gockeler:1996hg}.

There is also a possible mixing between $\cO_1$
and the lower-dimensional operator $\cO_8$~(\ref{O8}).
Indeed, we find in the one-loop approximation that the
vertex function of $\cO_1$ contains a term $\propto 1/a$:
\begin{equation}
  \cO_1\bigg|_{\frac{1}{a}-{\rm part}} =
  \gRcf (-0.5177)\,\frac{1}{a}\, \cO_8^{\rm Born}\,.
\end{equation}
This mixing leads to a contribution which diverges like
the inverse lattice spacing in the continuum limit.
The operator $\cO_8$ has to be subtracted non-perturbatively
from the operator $\cO_1$  which might be a
difficult task in simulations.

\subsubsection{$\cO^{5,DD}_{\{124\}} \, (\tau_3^{(4)},C=+1)$}

For this representation we find mixing between $\cO^{5,DD}_{\{124\}}$ and
$\cO^{5,\partial\partial}_{\{124\}}$.
The anomalous dimension matrix is given by (\ref{andim1}), the finite
contributions are collected in the matrix
\begin{equation}
  c_{jk}^{(I,II)} =\left( \begin{array}{rr}
    -12.1171     & 0.1667\\
      0          & 15.7963
  \end{array}
  \right)  \,.
\end{equation}
As in the case of the operator $\cO_{\{124\}}^{DD}$
above, the mixing is rather small, and the diagonal matrix elements
$c_{11}$ and $c_{22}$ can be inferred from previous
work~\cite{Gockeler:1996hg}.

\subsubsection{$\cO_1^5 \, ( \tau_2^{(8)},C=+1)$}

First we discuss the mixing of operators of the same dimension.
The set of contributing operators
is found to be
$$
  \{\cO^5_1,\cO^5_2,\cO^5_3,\cO^5_4,\cO^5_5,\cO^5_6 \} \,.
$$
As in the case of the operator $\cO_1$, one operator -- here $\cO^5_7$ --
does not contribute to mixing in one-loop order.
The finite contributions for the two lattice derivatives are
\begin{equation}
  c_{jk}^{(I)} =\left( \begin{array}{rrrrrr}
  -12.8609     & -2.0653      &  0.3490     &  0.8538     & 0.0511  & 0.0594\\
    0          &  15.7963     & 0           & 0           & 0       & 0\\
  3.4220       &  15.8207     & -15.3592    & -5.1639     &  0.1701 & -0.9431 \\
    0          & 0            & 0           & 15.7963     & 0       & 0 \\
    0          & -8.9124      & 0           & 0           & 0.9597  & -0.9597 \\
    0          & -8.9124      & 0           & 0           & -0.3199  & 0.3199
  \end{array}
  \right)
\end{equation}
and
\begin{equation}
  c_{jk}^{(II)} =\left( \begin{array}{rrrrrr}
  -12.8609     & 1.4894       & 0.3490     & -0.3311      & 0.0511  & 0.0594\\
    0          &  15.7963     & 0           & 0           & 0       & 0\\
 3.4220        & -7.3002      & -15.3592    & 2.5431      & 0.1701  & -0.9431 \\
    0          & 0            & 0           & 15.7963     & 0       & 0 \\
    0          & -8.9124      & 0           & 0           & 0.9597  & -0.9597 \\
    0          & -8.9124      & 0           & 0           & -0.3199 & 0.3199
  \end{array}
  \right) \,.
\end{equation}
The anomalous dimension matrix is the same as for the operators without
$\gamma_5$, see (\ref{andim2}). Again, some of the mixing coefficients
may be non-negligible. The mixing between $\cO^5_1$ and $\cO^5_3$
is visible even in the forward case.

We also find mixing with a lower-dimensional operator, in this case it
is the operator $\cO_8^5$~(\ref{O58}). The corresponding contribution in
the vertex function of $\cO^5_1$ reads
\begin{equation}
  \cO_1^5\bigg|_{\frac{1}{a}-{\rm part}} =
  \gRcf (-0.2523)\,\frac{1}{a}\,\cO_8^{5,{\rm Born}} \,.
\end{equation}

\subsubsection{$\cO^T_1 \, (\tau_2^{(3)},\, C=-1)$}

For the mixing between the operators $\cO^T_1$ (\ref{OT1}) and
$\cO^T_2$ (\ref{OT2}) the matrix of anomalous dimensions is given by
\begin{equation}
  \gamma_{jk}=\left( \begin{array}{rr}
    \frac{13}{3} & -\frac{2}{3}\\
      0          & 1
  \end{array}
  \right) \,,
\label{andimT1}
\end{equation}
and the finite contributions are
\begin{equation}
  c_{jk}^{(I,II)} =\left( \begin{array}{rr}
    -11.5483     & 0.2189 \\
      0          & 17.0181
  \end{array}
  \right) \,.
  \label{cjkT1}
\end{equation}

\subsubsection{$\cO^T_3 \, (\tau_3^{(3)},\, C=-1)$}

The mixing between the operators $\cO^T_3$ (\ref{OT3}) and
$\cO^T_4$ (\ref{OT4}) leads to the same anomalous dimension matrix
(\ref{andimT1}) as the previous case. For the finite pieces we obtain
\begin{equation}
  c_{jk}^{(I,II)} =\left( \begin{array}{rr}
  -11.8688     & 0.2753 \\
    0          & 17.0181
  \end{array} \right) \,.
\label{cjkT2}
\end{equation}

\subsubsection{$\cO^T_5 \, (\tau_2^{(6)},\, C=-1)$}

For the operators $\cO^T_5$ (\ref{OT5}) and $\cO^T_6$ (\ref{OT6}) we find
the finite mixing contributions
\begin{equation}
  c_{jk}^{(I,II)} =\left( \begin{array}{rr}
  -11.7477     & 0.2380 \\
    0          & 17.0181
  \end{array}
  \right)
  \label{cjkT3}
\end{equation}
with the identical anomalous dimension matrix (\ref{andimT1}).

\section{Summary}

Within the framework of lattice QCD with Wilson fermions and Wilson's
plaquette action for the gauge fields
we have calculated the one-loop quark matrix elements of operators needed for
the second moments of GPDs and meson distribution amplitudes.
{}From these we have determined the matrices
of renormalisation and mixing coefficients in the $\MS$-scheme.
The operators contributing to the considered moments
in the continuum contain two derivatives and are hence of dimension 5.
On the lattice we should take operators with definite C-parity belonging
to multiplets which transform irreducibly under the hypercubic group.
We have selected some operators commonly used in numerical simulations
of ordinary parton distributions. While these distributions are
extracted from forward matrix elements, we have to consider non-forward
matrix elements in order to study GPDs. This circumstance enlarges the
set of potentially mixing operators, because also operators with external
ordinary derivatives have to be taken into account.

For the representations $(\tau_2^{(4)},C=-1)$ and $(\tau_3^{(4)},C=+1)$
there are only two multiplets of operators of dimension $\leq 5$. Actually,
they are all of dimension 5, and the mixing pattern is the same as in the
continuum. The mixing coefficients turn out to be quite small.

In the case of the representations $(\tau_1^{(8)},C=-1)$ and
$(\tau_2^{(8)},C=+1)$ we have eight multiplets of operators of dimension
$\leq 5$. In the one-loop approximation one of them cannot yet contribute,
but the remaining seven multiplets do actually show up. Six of them are of
dimension 5, but there is also one multiplet of operators of dimension 4.
For these representations some of the mixing coefficients are larger
than in the above cases. In particular the mixing with the lower-dimensional
operators, which will probably receive sizeable non-perturbative
corrections, might lead to difficulties in the numerical simulations:
$1/a$ effects are hard to get under control.

In the operators just mentioned the Dirac matrix is either $\gamma_\mu$ or
$\gamma_\mu \gamma_5$. In connection with (generalised) transversity
distributions operators with $[\gamma_\mu,\gamma_\nu]$ are of interest. Here we
restricted ourselves to cases where only mixing with the external
derivative counterparts is possible and found rather small mixing
coefficients.

The results presented in this paper were obtained for Wilson fermions.
However, numerical simulations will be performed with clover or
overlap fermions. Therefore, the numbers given in the preceding
section can only serve as first hints at the problems which will occur
in these calculations, especially the mixing problem. The sets of
mixing operators, on the other hand, follow from symmetry arguments
and are therefore universally valid, although additional symmetries
may lead to further restrictions, e.g.\ in the case of overlap fermions.
It should also be noted that the overall normalisations of the
operators are somewhat arbitrary. So some care has to be exercised
when using our results in concrete applications.

\section*{Acknowledgements}
We thank O. V. Tarasov for useful discussions. This work
is supported by DFG under contract FOR 465 (Forschergruppe
Gitter-Hadronen-Ph\"{a}nomeno\-logie).

\section*{Appendix: Feynman rules}

We need Feynman rules for the lattice operators (\ref{OpDD}), (\ref{Op5DD}) and
(\ref{Optrans}).
Denoting the momentum of the outgoing quark by $p'$, that of
the  incoming quark by $p$ and the incoming gluon momenta by $k_i$,
we use the following Fourier decomposition for the quark and gauge fields,
\begin{eqnarray}
  \psi(x)&=& \int
  \frac{d^4p}{(2 \pi)^4} \psi(p) \, {\rm e}^{{\rm i} p \cdot x}
  \,,\nonumber
  \\
  \bar{\psi}(x)&=& \int
  \frac{d^4p'}{(2 \pi)^4} \bar{\psi}(p') \, {\rm e}^{-{\rm i} p' \cdot  x}
  \,,
  \\
  A_\mu(x)&=& \int
  \frac{d^4k_i}{(2 \pi)^4} A_\mu(k_i)\, {\rm e}^{{\rm i} k_i
  \cdot  (x + a \hat{\mu}/2)} \,,
  \nonumber
\end{eqnarray}
where the momenta are restricted to the first Brillouin zone.

We use the standard realisation of the covariant derivatives
acting to the right and to the left
\begin{eqnarray}
  \Dr_\mu \psi (x)  &=& \frac{1}{2a} \Big[U_{x,\mu} \,
  \psi(x+a\hat{\mu}) -U^\dagger_{x-a\hat{\mu},\mu} \, \psi(x-a\hat{\mu}) \Big]
  \,, \nonumber
  \\
  \bar{\psi} (x) \Dl_\mu &=& \frac{1}{2a} \Big[
  \bar{\psi} (x+a\hat{\mu})\, U^\dagger_{x,\mu} - \bar{\psi}(x-a\hat{\mu})
  \, U_{x-a\hat{\mu},\mu} \Big] \,.
  \label{Dlr}
\end{eqnarray}
The external ordinary derivative is taken as
\begin{equation}
  \partial_\mu \left( \bar{\psi} \cdots \psi \right)\!(x)
  =
  \frac{1}{a}\left[\left(\bar{\psi} \cdots \psi \right) (x+a\hat{\mu})
  -                 \left(\bar{\psi} \cdots \psi \right)\!(x)
             \right] \,.
  \label{totderiv2}
\end{equation}

There are two convenient ways to define the corresponding operators in
momentum space with non-zero momentum transfer $q$.
One possibility is to ``act'' with $q$ at the lattice point $x$.
For the case of one covariant derivative this leads to
(setting the Dirac matrix in the operator equal to the unit matrix
for simplicity)
\begin{eqnarray}
  & &\hspace{-1cm}\left(\bar{\psi}\Dlr_\mu \psi\right)^{(I)}\!\!(q) = \sum_x
  \left(\bar{\psi}\Dlr_\mu \psi\right)\!
  (x)\,\, {\rm e}^{ i q \cdot x}
  \nonumber \\
  & & =
  \frac{1}{2a}\sum_x \left[\bar{\psi}(x)U_{x,\mu} \psi(x+a\hat{\mu})-
  \bar{\psi}(x+a\hat{\mu})U^\dagger_{x,\mu} \psi(x)\right]\,
   \left[{\rm e}^{i q\cdot  x}+{\rm e}^{i q \cdot(x+a\hat{\mu})}\right]\,.
  \label{DI}
\end{eqnarray}
Alternatively, the momentum transfer can be applied to the
``position center'' (for the case of one covariant derivative $\vec{D}$)
as follows
\begin{eqnarray}
  \left(\bar{\psi}\Dr_\mu \psi\right)(q) &=&
  \frac{1}{2a}\sum_x \Bigl[\bar{\psi}(x)U_{x,\mu} \psi(x+a\hat{\mu})\,
  \, {\rm e}^{ i q \cdot (x+{a\hat{\mu}}/{2})} \nonumber\\
   & &\quad  -\bar{\psi}(x) U^\dagger_{x-a\hat{\mu},\nu}\psi(x-a\hat{\mu})\,
  \, {\rm e }^{ i q \cdot (x-{a\hat{\mu}}/{2})}
  \Bigr] \,,
  \nonumber
\end{eqnarray}
which leads to
\begin{eqnarray}
  & &\hspace{-1cm}\left(\bar{\psi}\Dlr_\mu \psi\right)^{(II)}\!\!(q)
  \nonumber \\
  & & =
  \frac{1}{a}\sum_x \left[\bar{\psi}(x)U_{x,\mu} \psi(x+a\hat{\mu})-
  \bar{\psi}(x+a\hat{\mu})U^\dagger_{x,\mu} \psi(x)\right]\,
  {\rm e}^{i q   \cdot(x+a\hat{\mu}/2)}\, .
  \label{DII}
\end{eqnarray}
This realisation might be more suitable for numerical simulations.
Note that
\begin{equation}
  \left(\bar{\psi}\Dlr_\mu \psi\right)^{(I)}\!\!(q)=\cos \frac{a q_\mu}{2} \,
  \left(\bar{\psi}\Dlr_\mu \psi\right)^{(II)}\!\!(q) \,.
\end{equation}
For the external ordinary derivative we use
\begin{eqnarray}
  \partial_\mu \left(\bar{\psi}\cdots\psi\right)\!(q)
  & = & \frac{1}{a} \sum_x
      \left[ \left( \bar{\psi} \cdots \psi \right)(x+a\hat{\mu})
  - \left( \bar{\psi} \cdots \psi \right) (x) \right]
         {\rm e}^{i q \cdot (x+a\hat{\mu}/2)}
  \nonumber \\
  & = &  - \frac{2i}{a}
  \sin \frac{a q_\mu}{2} \, \left( \bar{\psi}\cdots \psi \right) \!(q) \,.
  \label{dII}
\end{eqnarray}
The corresponding Feynman rules up to
$O(g^2)$ derived from expressions based on (\ref{DI})
and (\ref{DII})
are marked by the superscripts (I) and (II), respectively.

\underline{$O(g^0)$}
\begin{eqnarray}
  \cO^{(II),DD}_{\mu\nu\omega}&=& \bar{\psi}(p')\gamma_\mu\psi(p)\,
  \frac{1}{a^2} \sin \frac{a( p+ p')_\nu   }{2}
  \sin \frac{a( p+ p')_\omega}{2}
  \,,
  \nonumber
  \\
  \cO^{(I),DD}_{\mu\nu\omega}&=&
  \cos \frac{a (p-p')_\nu }{2}
  \cos \frac{a (p-p')_\omega }{2} \,
  \cO^{(II),DD}_{\mu\nu\omega}
  \,,
  \nonumber
  \\
  \cO^{(II),\partial D}_{\mu\nu\omega}&=&
  \bar{\psi}(p')\gamma_\mu\psi(p) \,\frac{1}{a^2}
  \sin \frac{a( p- p')_\nu   }{2}
  \sin \frac{a( p+ p')_\omega}{2}
  \,,
  \\
  \cO^{(I),\partial D}_{\mu\nu\omega}&=&
  \cos \frac{a (p-p')_\omega }{2} \,
  \cO^{(II),\partial D}_{\mu\nu\omega}
  \,,
  \nonumber
  \\
  \cO^{\partial \partial}_{\mu\nu\omega}&=&
  \bar{\psi}(p')\gamma_\mu\psi(p) \,\frac{1}{a^2}
  \sin \frac{a(p-p')_\nu}{2}
  \sin \frac{a(p-p')_\omega}{2}
  \,.
  \nonumber
\end{eqnarray}

\underline{$O(g)$}
\begin{eqnarray}
  \cO^{(II),DD}_{\mu\nu\omega}&=& g
  \sum_\sigma \bar{\psi}(p')\gamma_\mu A_\sigma(k_1) \psi(p)
  \,
  \cos\frac{a (p+p')_\sigma}{2}
  \nonumber
  \\
  &&
  \times
  \frac{1}{a}
  \left[
  \delta_{\nu\sigma}    \sin\frac{a (p+p'-k_1)_\omega}{2}
  +
  \delta_{\omega\sigma} \sin\frac{a (p+p'+k_1)_\nu}   {2}
  \right]
  \nonumber
  \,,
  \\
  \cO^{(I),DD}_{\mu\nu\omega}&=&
  \cos \frac{a (p-p'+k_1)_\nu }{2}
  \cos \frac{a (p-p'+k_1)_\omega }{2} \,
  \cO^{(II),DD}_{\mu\nu\omega}
  \,,
  \\
  \cO^{(II),\partial D}_{\mu\nu\omega}&=&
  g \,\bar{\psi}(p')\gamma_\mu A_\omega(k_1) \psi(p)
  \,
  \frac{1}{a} \sin\frac{a (p-p'+k_1)_\nu}{2}
  \cos\frac{a (p+p')_\omega }{2}
  \,.
  \nonumber
  \\
  \cO^{(I),\partial D}_{\mu\nu\omega}&=&
  \cos \frac{a (p-p'+k_1)_\omega }{2} \,
  \cO^{(II),\partial D}_{\mu\nu\omega}
  \,.
  \nonumber
\end{eqnarray}

\underline{$O(g^2)$ for the tadpole case $k_2=-k_1$}

Here we restrict ourselves to the tadpole case $k_2=-k_1$
(cf.\ Fig.~\ref{fig:vertex}) as this is all we need in this paper.
For general gluon momenta the Feynman rules are much more complicated.
\begin{eqnarray}
  \cO^{(II),DD}_{\mu\nu\omega}&=&
  \frac{g^2}{2}\sum_{\sigma,\tau} \bar{\psi}(p')\, \gamma_\mu \,
  A_\tau(-k_1)   A_\sigma(k_1) \psi(p)
  \nonumber
  \\
  &&
  \times
  \Biggl\{
  2 \delta_{\tau\nu} \delta_{\sigma\omega}
   \cos \frac{a (p+p'+k_1)_\nu   }{2}
   \cos \frac{a (p+p'+k_1)_\omega}{2}
  \nonumber
  \\
  &&
  -\left(\delta_{\tau\nu} \delta_{\sigma\nu} + \delta_{\tau\omega}
  \delta_{\sigma\omega}  \right)
  \sin \frac{a (p+p')_\nu   }{2}
  \sin \frac{a (p+p')_\omega}{2}
  \Biggr\}
  \,,
  \nonumber
  \\
  \cO^{(I),DD}_{\mu\nu\omega}&=&
  \cos \frac{a (p-p')_\nu }{2}
  \cos \frac{a (p-p')_\omega }{2} \,
  \cO^{(II),DD}_{\mu\nu\omega}
  \,,
  \\
  \cO^{(II),\partial D}_{\mu\nu\omega}&=&
  -\frac{g^2}{2}
  \bar{\psi}(p')\, \gamma_\mu \, A_\omega(k_1) A_\omega(-k_1) \psi(p)
  \,
  \sin\frac{a (p-p')_\nu}   {2}
  \sin\frac{a (p+p')_\omega}{2}
  \,,
  \nonumber
  \\
  \cO^{(I),\partial D}_{\mu\nu\omega}&=&
  \cos \frac{a (p-p')_\omega }{2} \,
  \cO^{(II),\partial D}_{\mu\nu\omega}
  \,.
  \nonumber
\end{eqnarray}
The Feynman rules for the operators with different Dirac matrices
are obtained by the obvious replacements.
Note that the forward case is realised
for $O(g^0)$ and $O(g^2)$ (in the tadpole case) by $p=p'$,
for $O(g)$ we have to take $p'=p+k_1$.

\end{document}